\documentclass[useAMS,usenatbib,usedcolumn]{mn2e}

\def\HI{\ifmmode{\rm HI}\else{H\/{\sc i}}\fi}

\def\lsun{\ifmmode{{\mathrm L}_{\odot}}\else{L$_{\odot}$}\fi} 

\def\deg{\hbox{$^\circ$}}

\def\msun{\ifmmode{{\mathrm M}_{\odot}}\else{M$_{\odot}$}\fi} 
\def\msunpc2{\ifmmode{{\mathrm M}_{\odot} \, {\mathrm{pc}}^{-2}}\else{M$_{\odot} \, {\mathrm {pc}}^{-2}$}\fi}

\usepackage[figuresright]{rotating}
\usepackage{lscape}
\usepackage{graphics}
\usepackage{epsfig}
\usepackage{multirow}
\usepackage{bigdelim}
\usepackage{bigstrut}

\defcitealias{Noordermeer05}{N05}
\defcitealias{Noordermeer07b}{N07}

\title[MOND and early-type disc galaxies]{Confrontation of MOND with the
  rotation curves of early-type disc galaxies}
\author[R.H.~Sanders \& E.~Noordermeer]{R.H.~Sanders$^1$\thanks{email:
  sanders@astro.rug.nl} and E.~Noordermeer$^2$ \\  
  $^1$Kapteyn Astronomical Institute, University of Groningen, P.O.~Box 800,
  9700 AV Groningen, The Netherlands \\ 
  $^2$University of Nottingham, School of Physics and Astronomy, University
  Park, NG7 2RD Nottingham, UK}
\begin{document}

\date{submitted: 13-03-2007}

\maketitle

\begin{abstract}
We extend the MOND analysis to a sample of 17 high surface brightness, 
early-type disc galaxies with rotation curves derived from a combination of 
21cm \HI\ line observations and optical spectroscopic data. A number of
these galaxies have asymptotic rotation velocities between 250 and 350~km/s
making them among the most massive systems (in terms of baryonic mass)
considered in the context of MOND. We find that the general MOND prediction
for such galaxies -- a rotation curve which gradually declines to the
asymptotic value -- is confirmed, and in most cases the MOND rotation curve,
determined from the mean radial light and gas distribution, agrees in detail
with the observed rotation curve. In the few cases where MOND appears not to
work well, the discrepancies can generally be understood in terms of various
observational errors -- such as incorrect orientation angles and/or distances
-- or of unmodelled physical effects -- such as non-circular motions.

The implied mass-to-light ratios for the stellar disc and bulge constrain the
MOND interpolating function; the form recently suggested by Zhao \& Famaey
(2005) yields more sensible values than the one traditionally used in MOND
determinations of galaxy rotation curves.   
\end{abstract}


\section{Introduction}
\label{sec:Introduction}
As an acceleration-based modification of Newtonian dynamics, MOND
makes general predictions about the contrasting form of rotation
curves in low surface brightness (LSB) and high surface brightness
(HSB) galaxies \citep{Milgrom83}. If surface density is roughly proportional
to surface brightness, then LSB galaxies have low internal accelerations and
MOND predicts a large discrepancy between the visible and Newtonian dynamical
mass within the optical image; in particular, in these systems the rotation
curves should slowly rise to the final asymptotic circular velocity. On the
other hand, HSB galaxies, within the bright inner regions, are in the high
acceleration, or Newtonian, regime.  MOND predicts that there should be 
a small discrepancy between
the observable mass and the dynamical mass within the bright optical image;
moreover, the rotation curves should rapidly rise and then fall in
an almost Keplerian fashion to the final asymptotic value.  
\HI\ observations of several galaxies
spanning a range of surface brightness certainly exhibit this general trend
\citep{Casertano91} but the sample is small and the data are inhomogeneous.  

With respect to LSB galaxies the MOND predictions have been spectacularly
confirmed by \HI\ observations of such objects over the past decade
\citep{McGaugh98b, Sanders98}. Most recently, \citet{Milgrom06} have
extended this analysis to several galaxies with extremely low mass or rotation
velocity -- systems where the observed distribution of neutral gas dominates
the detectable mass budget of the galaxies and the mass-to-light ratio of the 
stellar component effectively disappears as a free fitting parameter. 
They demonstrated that MOND generally accounts for the form and the amplitude
of rotation curves in these systems.  

At the other extreme, there has not been much consideration of the high
luminosity or HSB galaxies.  The gravitational force distribution
probed by planetary nebulae kinematics in several elliptical
galaxies \citep{Romanowsky03} is consistent with MOND \citep{Milgrom03} --
primarily the basic prediction of a small discrepancy in high surface
brightness systems. However, because of the uncertain degree of anisotropy in
the velocity distribution of these probes, rotation curves remain a far more
unambiguous tracer of the radial dependence of force.

A sample of eight rapidly rotating, but late-type, disc galaxies has been
presented by \citet{Spekkens06}, but these objects are apparently not as  
centrally concentrated or bulge-dominated as the early type rapid
rotators that we require in order to confront the generic MOND predictions. 
The rotation curves of several early type disc galaxies in the published
literature at the time were considered in terms of MOND by \citet{Sanders96},
but the data were quite inhomogeneous and the derived rotation curves were not
of the same quality as in more gas rich systems. The basic problem is that
high luminosity, HSB galaxies tend to be of early-type (S0-Sb) and have
generally low gas content with patchy distributions.

Recently, a more homogeneous sample of early-type disc galaxies with
high-quality \HI\ observations has become available \citep[][hereafter
N05]{Noordermeer05}. In \citet[][hereafter N07]{Noordermeer07b}, these data
were combined with high-resolution optical spectroscopic observations to
derive rotation curves for 19 systems. These galaxies have morphological types
ranging from S0 to Sab, and have, for the most part, an apparent spheroidal
bulge component and a high central surface brightness.   
They are mostly massive systems, with maximum rotation velocities of 200 -- 500
km/s. Significantly, surface photometry in the B- and R-bands exists for all
these galaxies \citep{Noordermeer07a}; this is necessary to estimate
the distribution of visible matter and the resulting Newtonian acceleration. 

In this paper, we extend the MOND analysis to this sample of HSB, rapidly
rotating galaxies. The rotation curves follow the general MOND predictions. 
In the bright inner regions there is little discrepancy between the
observed and Newtonian rotation curves, while at larger radii the rotation
curves often decline to an asymptotically constant value. 
Applying the MOND formula to the mass distribution as traced by the
visible light and \HI\ emission, we find that in most cases the observed
rotation curve is well matched by the MOND prediction. There are a few
exceptions where the observed rotation curves appear to deviate from the MOND
predictions, and these cases are considered in more detail.  

Because the accelerations range smoothly from the Newtonian to the MOND
regime, this sample of rotation curves can be used to constrain the form of
the MOND interpolating function $\mu$.  Using the typically applied form for
this function we find that the mass-to-light
ratios of the bulge and disc are often implausible; in particular, in several
cases a higher M/L is required for the disc than for the bulge. Using the
simpler interpolating function recently suggested by 
\citet[]{Zhao06}, the M/L values become more
reasonable with the disc values generally comparable to or lower than those of
the bulge. 

The remainder of this paper is structured as follows. In
section~\ref{sec:sample+data}, we describe the sample of galaxies and the 
observational data. Section~\ref{sec:MONDrcs} describes the procedures we used
to derive the predicted MOND rotation curves. In section~\ref{sec:results}, we
present the general results, while individual galaxies are discussed in
section~\ref{sec:individualgalaxies}. Finally, in
section~\ref{sec:conclusions}, we draw some conclusions. 
\begin{table*}
 \begin{minipage}{15.6cm}
  \centering
   \caption[Rotation curve sample galaxies: basic data]
   {Sample galaxies: basic data. (1)~UGC number; (2)~alternative name;
    (3)~morphological type; (4)~distance (based on Hubble flow, corrected for
    Virgo-centric inflow and assuming h=0.75); (5)~and (6)~B- and R-band
    luminosities (corrected for Galactic foreground extinction); (7)~central
    surface brightness (R-band; corrected for Galactic foreground extinction);
    (8)~maximum and (9)~asymptotic rotation velocity and (10)~centripetal 
    acceleration at last point in rotation curve. Column (3) was taken from
    NED, (4) from \citet{Noordermeer05} and (5) -- (7) from  
    \citet{Noordermeer07a}. \label{table:sample}}   
  
   \begin{tabular}{r@{\hspace{0.6cm}}llccccccc}
    \hline 
    \multicolumn{1}{c@{\hspace{0.6cm}}}{UGC} &
    \multicolumn{1}{c}{\hspace{-0.2cm}alternative} & 
    \multicolumn{1}{c}{Type} & D & L$_{\mathrm B}$ & 
    L$_{\mathrm R}$ & $\mu_{0,R}$ & $V_m$ & $V_f$ & $a_f$ \\     
    
    & \multicolumn{1}{c}{\hspace{-0.2cm}name} &
    & Mpc & $10^{10}$ L$_\odot$ & $10^{10}$ L$_\odot$ &
    $\frac{\mathrm{mag}}{\mathrm{arcsec}^2}$ & km s$^{-1}$ & km s$^{-1}$ &
    $10^{-8}$ cm s$^{-2}$ \\   
   
    \multicolumn{1}{c@{\hspace{0.6cm}}}{(1)} &
    \multicolumn{1}{c}{\hspace{-0.2cm}(2)} & \multicolumn{1}{c}{(3)} & 
    (4) & (5) & (6) & (7) & (8) & (9) & (10) \\  
    \hline 

    2487  & NGC 1167 & SA0-         & 67.4 & 8.79           & 10.47 & 17.00 & 390 & 320 & 0.46 \\
    2916  & --       & Sab          & 63.5 & 4.09           & 3.37  & 16.97 & 220 & 200 & 0.29 \\
    2953  & IC 356   & SA(s)ab pec  & 15.1 & 4.79           & 5.50  & 15.72 & 315 & 260 & 0.44 \\
    3205  & --       & Sab          & 48.7 & 3.53           & 2.99  & 17.17 & 240 & 210 & 0.40 \\
    3546  & NGC 2273 & SB(r)a       & 27.3 & 1.59           & 1.84  & 15.84 & 260 & 190 & 0.43 \\
    3580  & --       & SA(s)a pec:  & 19.2 & 0.328          & 0.311 & 17.87 & 127 & 120 & 0.20 \\
    3993  & --       & S0?          & 61.9 & 1.85           & 1.84  & 16.99 & 300 & 250 & 0.41 \\
    4458  & NGC 2599 & SAa          & 64.2 & 5.55           & 5.86  & 16.00 & 490 & 260 & 0.34 \\
    5253  & NGC 2985 & (R')SA(rs)ab & 21.1 & 3.44           & 3.05  & 15.56 & 255 & 220 & 0.31 \\
    6786  & NGC 3900 & SA(r)0+      & 25.9 & 1.47$^\dagger$ & 1.50  & 15.88 & 230 & 210 & 0.48 \\
    6787  & NGC 3898 & SA(s)ab      & 18.9 & 1.56           & 1.72  & 15.35 & 270 & 250 & 0.64 \\
    8699  & NGC 5289 & (R)SABab:    & 36.7 & 0.964          & 1.05  & 16.47 & 205 & 180 & 0.45 \\
    9133  & NGC 5533 & SA(rs)ab     & 54.3 & 4.79           & 5.92  & 16.27 & 300 & 230 & 0.17 \\
    11670 & NGC 7013 & SA(r)0/a     & 12.7 & 0.745          & 0.879 & 15.59 & 190 & 160 & 0.40 \\
    11852 & --       & SBa?         & 80.0 & 2.33           & 2.17  & 17.50 & 220 & 160 & 0.09 \\
    11914 & NGC 7217 & (R)SA(r)ab   & 14.9 & 2.00           & 1.84  & 16.05 & 305 & 290 & 3.48 \\
    12043 & NGC 7286 & S0/a         & 15.4 & 0.160          & 0.107 & 18.71 & 93  & 85  & 0.18 \\
    \hline                                                                                   
    \multicolumn{10}{l}{$^\dagger$ No B-band data available in \citet{Noordermeer07a};
                       L$_{\mathrm B}$ taken from LEDA.}   
   \end{tabular}
 \end{minipage}
\end{table*}  

\section{The sample and observational data}
\label{sec:sample+data}
The rotation curves in \citetalias{Noordermeer07b} are based on the \HI\ data
from \citetalias{Noordermeer05}, which in turn were obtained in the framework
of the Westerbork survey of \HI\ in spiral and irregular galaxies 
\citep[WHISP;][]{Kamphuis96, vanderHulst01}. A full description of the sample
selection is given in \citetalias{Noordermeer07b}; here we briefly mention a
few points relevant for this study. 

All galaxies in \citetalias{Noordermeer07b} have well-resolved \HI\ velocity
fields (at least 5 to 10 independent beams across), and sufficient \HI\ across
the disc to define a complete two-dimensional field. Moreover, their velocity
fields show no obvious evidence for large-scale deviations from circular
motions about the centre of the galaxy: no strongly interacting or strongly
barred galaxies were selected to avoid complications introduced by
non-circular motions.  

Of the 19 galaxies in N07, two have been dropped here. The first (UGC~624) has
an asymmetric velocity field, undermining the quality of its rotation curve. A
second (UGC~4605) is nearly edge-on and no photometry is available to trace
the stellar mass distribution. 

The observational properties of the remaining sample of 17 galaxies is given
in Table 1. 
Note that all distances are estimated from the systemic velocity, corrected
for Virgo-centric inflow and assuming h=0.75.

For all of these objects, the \HI\ velocity field has been supplemented by
higher resolution single slit optical spectra in the central regions. This
is evident as a much higher density of data points defining the inner observed 
rotation curves. While higher resolution data is certainly desirable in the
region where the rotation velocity is changing rapidly, a note of caution is
necessary. These data points are based upon single slit spectra and the
implicit assumption is that the observed velocities along the slit can
directly be translated into rotational velocities (using the position angle
and inclination as defined by the \HI\ velocity field). But streaming motions
of gas, for example, by weak non-axisymmetric distortions in the central
regions, can introduce uncertainties in the interpretation of optical velocities
as circular velocities. 
Moreover, in any blind fitting procedure, the more numerous optical data
points should not be given weight equal the outer \HI\ points, to avoid that
the fitted model is determined entirely by the optical data. 

In addition to these considerations, there are the usual uncertainties which
complicate the interpretation of the rotation curves as tracers of the radial
force distribution. There are, in a number of cases, at least weak
non-axisymmetric distortions (e.g.\ bars) in the inner region that introduce
errors in the derived rotation curve. For other objects, the gas layer in the
outer regions is clearly warped. In principle, these effects were taken into
account as well as possible by using a tilted-ring model in the derivation of
the rotation curves \citep{Begeman87, Begeman89}: a set of rings each with its
own inclination, position angle and rotation velocity was fit to the radial
velocity data (see \citetalias{Noordermeer07b} for details). But this
procedure contains implicit physical assumptions -- such as
constant azimuthal velocity in a ring -- which may or may not be realistic. 
The overall consequence of these effects is that the presented rotation curves
cannot with certainty be taken as a true indicator of the force. The indicated
error bars are an attempt to quantify these uncertainties, but we cannot
exclude the possibility that in individual cases, the true circular velocity
deviates from the observed rotation curves. 

Finally, we use the photometric data from \citet{Noordermeer07a} to derive the
contribution to the gravitational field from the stars. Most of these galaxies
show evidence for two distinct components in the light distribution -- a
central spheroidal bulge and a more extended disc component. 
\citeauthor{Noordermeer07a} presented bulge-disc decompositions, where a bulge 
with a generalised S\'ersic profile was subtracted from the total light
distribution to give the remainder as the disc component. Assuming that the
disc is a flat system, with a vertical scale height one fifth of the radial
scale length, and comparing the ellipticity of the isophotes in the central
bulge-dominated regions with those of the outer disc, they could also estimate
the intrinsic axial ratio of the bulges.
Note that the bulge-disc decomposition introduces additional uncertainty in
our mass models. Usually, a range of bulge and disc parameters can be derived
to fit the observed intensity distribution. We have verified that, as long as
the bulge and disc mass-to-light ratios are comparable, differences in the
bulge-disc decompositions have little effect on the mass models. However, when
the mass-to-light ratios differ strongly, this is no longer the case, and
errors in the decomposition propagate into the mass models.

\section{Predicted MOND rotation curves}
\label{sec:MONDrcs}
The procedure here for generating a MOND rotation curve is basically
the same as described previously \citep[e.g.][]{Sanders02}.
One complication for this sample is the presence of a bulge which
may have a different mass-to-light ratio than the disc component.  This
adds uncertainty due to the decomposition procedure as well as
a second free parameter.

Basically one takes the photometry in each of the two components, bulge
and disc, as a faithful tracer of the stellar mass distribution
(mass-to-light ratio in each component does not vary with radius). Then given
the estimated intrinsic shape of the bulge and assuming that the disc is a
flat system, one may calculate, using the usual Poisson equation, the radial 
distribution of force (assuming a M/L ratio for each component). This
procedure is described in detail in \citet{Edosthesis}.
The observed surface density of neutral hydrogen is included (multiplying
by a factor of 1.3 to account for primordial helium) assuming this 
component to be in a disc with thickness equal to that of the stellar disc.

The Newtonian force $g_N$ is then converted into a ``true force'', $g$
by using the MOND formula
$$g\mu(g/a_0)= g_N.\eqno(1) $$
Here $a_0$ is the critical acceleration (the one new parameter of the
theory), which we fix at $1 \times 10^{-8} {\mathrm{cm/s^2}}$ 
\citep{Bottema02b}. $\mu$ is the MOND interpolating function which must
have the asymptotic dependence $\mu(x)\rightarrow x$ when $x\ll1$ (the MOND
limit) and $\mu(x)\rightarrow 1$ when $x\gg1$ (the Newtonian limit). Unlike
the case of the LSB galaxies, where $|g|/a_0<1$ everywhere, in these objects
there is a gradual transition from the Newtonian limit in the inner regions to
the MOND limit in the outer parts.  Therefore, one might expect the form of
$\mu$ to play some role in the fitted models, at least in the derived M/L
values. Two different forms have been considered in the recent literature.
Most often one takes
$$\mu(x) = {{x}\over \sqrt{1+x^2}},\eqno(2)$$
but more recently \citet{Zhao06} have suggested a simpler form which is
consistent with a multi-field relativistic extension of MOND
\citep{Bekenstein04}: 
$$\mu(x) = {x\over{1+x}}.\eqno(3)$$
\citet{Famaey06} have demonstrated that both functions produce reasonable fits
to observed rotation curves, but the second (eq.~3) generally yields lower
mass-to-light ratios of the visible components.  Here we will apply both forms
and consider the plausibility of the inferred M/L ratios as a constraint on
$\mu$. 

The circular velocity is determined from the true force ($v^2/r=g$)
and the M/L ratios of the two components are adjusted to achieve
the optimal fit to the observed curve. Thus, for the objects in this sample,
where there is a clear bulge component, there are two free parameters in the
fitting procedure. In view of the uncertainties associated with the errors on
the observed rotation curves (see previous section), a simple
$\chi^2$-minimisation does not always result in the best possible
fit. Instead, we have used the $\chi^2$-statistics only as a starting point
for inspection by eye of the individual fits, and in some cases adjusted the
mass-to-light ratios by hand to achieve a better fit to the overall shape of
the rotation curve.
\begin{figure*}
 \centerline{\psfig{figure=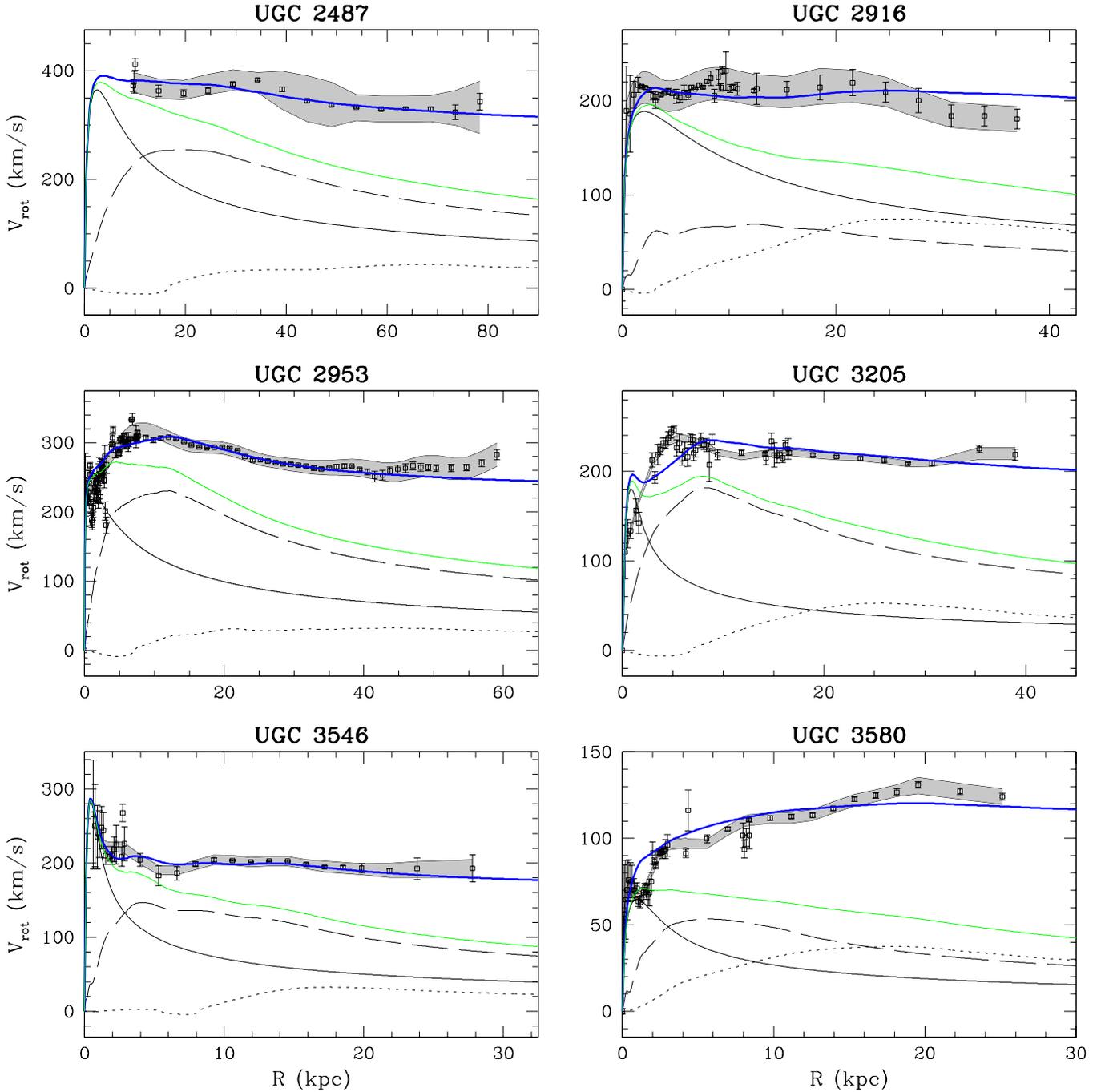,height=18.cm}}
 \caption{MOND fits to the observed rotation curves of early-type disc
   galaxies. Data points show the observed rotation curves and error bars give
   the combined uncertainties due to the measurement errors and kinematical
   asymmetries. The grey shaded bands give the allowed range due to
   inclination uncertainties. Thin black solid, dashed and dotted lines give
   the contributions from stellar bulge, disc and gas respectively. The thin
   green (grey) line gives the Newtonian sum of the individual components and
   the bold blue (grey) lines gives the total MOND rotation curve, assuming
   the interpolating function $\mu$ from eq.~3. 
 \label{fig:fits}}
\end{figure*}
\begin{figure*}
 \centerline{\psfig{figure=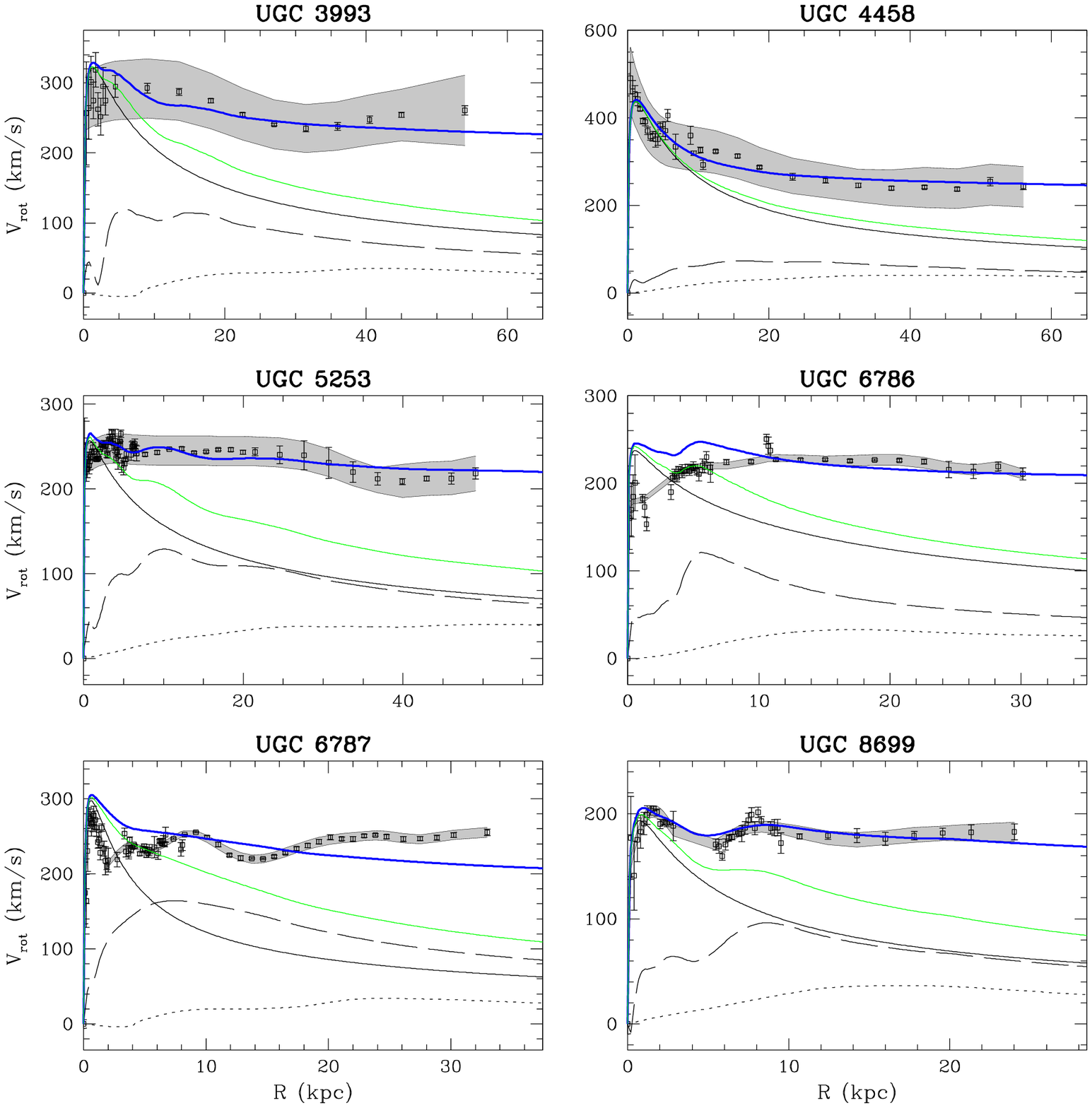,height=18.cm}}
 \contcaption{}
\end{figure*}
\begin{figure*}
 \centerline{\psfig{figure=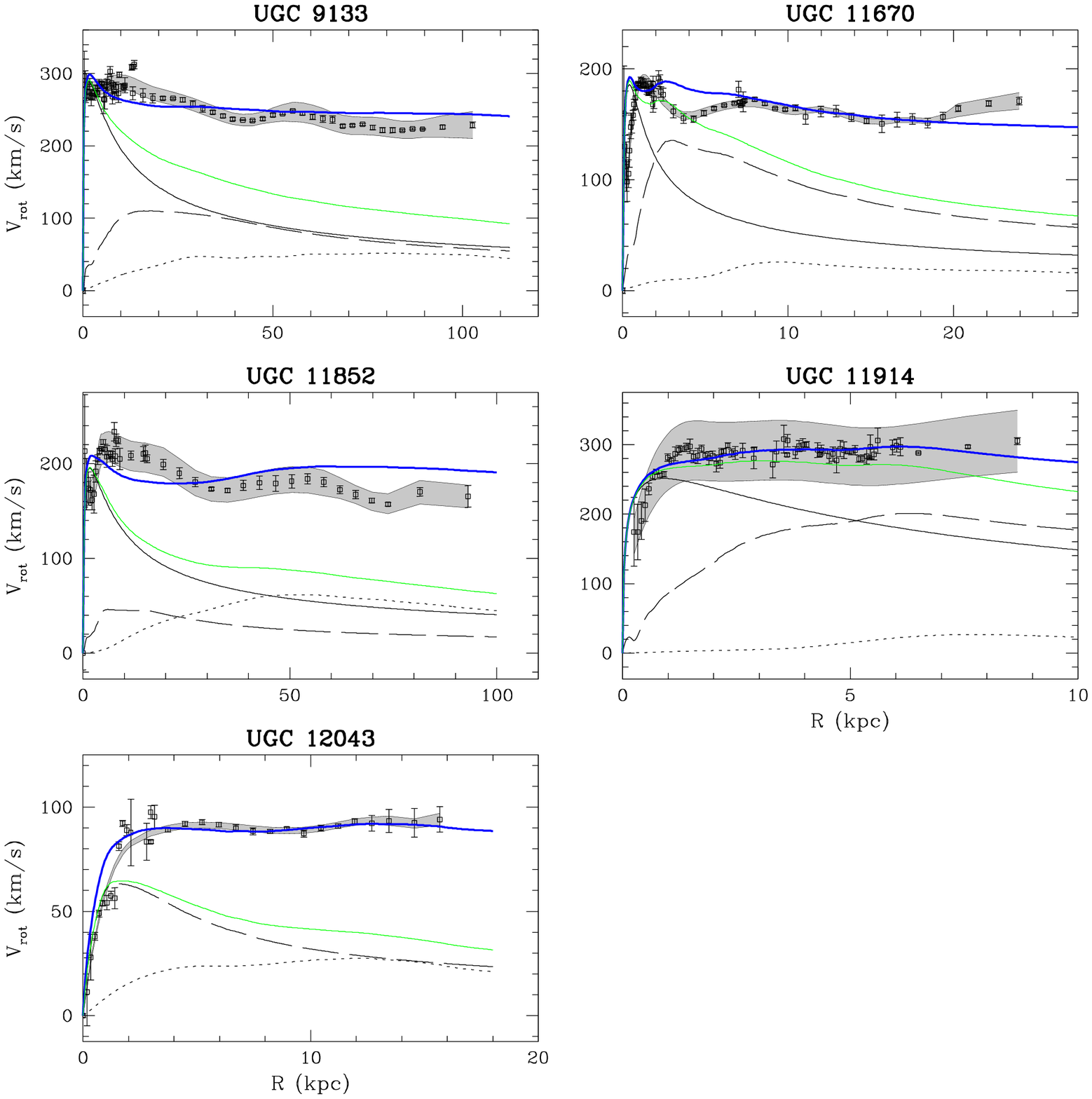,height=18.cm}}
 \contcaption{}
\end{figure*}

Finally, it should be noted that all distances were taken to be the Hubble
law values given by \citetalias{Noordermeer05} (assuming h=0.75 and corrected
for Virgo-centric inflow). Because most of these galaxies are quite distant,
Hubble expansion is generally a fair indicator, and distance was not allowed
to be an additional free parameter in these fits. However, a few galaxies lie
more nearby, in which case the Hubble distance can be off by a considerable
amount. In such cases, the quality of the fit might be improved further by
allowing the distance to vary within reasonable limits (see also
section~\ref{sec:individualgalaxies}).

\section{Results}
\label{sec:results}
The resulting MOND rotation curves are compared to the observed rotation
curves in figure~\ref{fig:fits}. Here the points with error bars show the
observed velocities. The thin, black solid line indicates the Newtonian
rotation curve of the bulge component, while the dashed and dotted curves show
the contributions of the stellar and gaseous discs respectively. These
Newtonian curves are, in form, those given by \citet{Edosthesis}.
The bold, blue solid curve is the resulting MOND rotation curve assuming the
interpolating function given by eq.~3. The implied mass-to-light ratios
of the two stellar components are given in Table 2.

\begin{table*}
 \begin{minipage}{10.5cm}
  \centering
   \caption{Fitted mass-to-light ratios in the R-band for discs and bulges
     separately, and for the entire galaxies. Columns 2 -- 4 refer to the fits
     with the `standard' function $\mu(x) = x / \sqrt{1 + x^2}$, whereas
     columns 5 -- 7 are for the form suggested by \citet{Zhao06}: 
     $\mu(x) = x / (1 + x)$. \label{t2}} 
   \begin{tabular}{r@{\hspace{0.9cm}}ccc@{\hspace{0.9cm}}ccc}
    \hline 
     & \multicolumn{3}{c@{\hspace{0.9cm}}}{\hspace{-0.2cm}--------- standard $\mu$ ---------} & 
    \multicolumn{3}{c}{\hspace{-0.2cm}-------------- $\mu_{\mathrm {ZF}}$ --------------} \\
    UGC & $({\mathrm{M/L}})_d$ & $({\mathrm{M/L}})_b$ & $({\mathrm{M/L}})_t$ & 
    $({\mathrm{M/L}})_d$ & $({\mathrm{M/L}})_b$ & $({\mathrm{M/L}})_t$ \\ 
    \multicolumn{1}{c@{\hspace{0.9cm}}}{(1)} & (2) & (3) & (4) & (5) & (6) & (7) \\ 
    \hline
    2487  &  6.0  & 9.0 &  6.7 & 4.6 & 6.5 & 5.1 \\
    2916  &  3.4  & 2.6 &  3.0 & 1.0 & 2.5 & 1.8 \\
    2953  &  5.3  & 3.4 &  5.0 & 3.5 & 5.0 & 3.8 \\
    3205  &  4.0  & 3.6 &  4.0 & 2.7 & 5.0 & 2.8 \\
    3546  &  4.0  & 5.9 &  4.2 & 2.5 & 5.9 & 2.9 \\
    3580  &  4.0  & 1.0 &  2.9 & 2.5 & 1.5 & 2.1 \\
    3993  & 26.   & 6.3 & 12.2 & 8.3 & 8.3 & 8.3 \\
    4458  &  2.5  & 7.0 &  4.6 & 1.0 & 6.0 & 3.4 \\
    5253  &  7.0  & 4.1 &  5.2 & 4.5 & 3.5 & 3.9 \\
    6786  & 10.8  & 4.3 &  5.6 & 5.5 & 6.0 & 5.9 \\
    6787  &  9.0  & 5.0 &  7.4 & 6.0 & 5.0 & 5.6 \\
    8699  &  8.5  & 4.1 &  5.9 & 4.6 & 4.0 & 4.2 \\
    9133  &  0.75 & 4.5 &  2.3 & 2.3 & 3.9 & 3.0 \\
    11670 &  4.0  & 3.4 &  3.9 & 3.0 & 3.5 & 3.1 \\
    11852 &  2.3  & 5.0 &  3.3 & 0.5 & 5.0 & 2.1 \\
    11914 &  8.8  & 6.8 &  7.9 & 6.8 & 6.8 & 6.8 \\
    12043 &  2.5  & --  &  2.5 & 2.2 & --  & 2.2 \\
    \hline
   \end{tabular}
 \end{minipage}
\end{table*}  

The error bars shown here require a word of explanation and
a note of caution.  There are three sources for the indicated
errors:  

First are the formal velocity fitting errors. For the velocities from the
optical spectra, these simply come from the derivation of the centre of the
emission line. For the velocities from the \HI\ data, they are given by the
non-linear least square routine that fits tilted rings to the entire
two-dimensional velocity fields \citep{Begeman87, Begeman89}. In the tilted
ring method it is assumed that deviations from planar circular motion can be 
modelled as pure rotation in a ring with a different inclination or position
angle on the sky.  While this may be a fair assumption in the outer regions
where warps are often seen, it is most likely not correct in the inner regions
where systematic deviations in the velocity field may result from non-circular
motion due to a non-axisymmetric gravitational field. In such cases, the formal
errors underestimate the true uncertainties. 

Secondly, the tilted ring analysis was performed separately for the
approaching and receding sides of the galaxy and this often resulted in
different rotation curves for the two different sides. This is also an
indication of asymmetries in the velocity fields, but not the sort of
bi-symmetric asymmetries that would be expected from bar-like distortions. The
error on the rotation curve due to these effects is taken to be one-fourth the
difference in the rotation curves of the two sides. 

Thirdly, there is an error resulting from the uncertainty in the inclination. 
This was estimated by eye looking at the scatter in the fitted inclination
resulting from the tilted ring program, and is most often the largest
contribution to the error budget.

The first two sources of errors were added quadratically to produce
the error bars shown in figure~\ref{fig:fits}. 
The latter uncertainty is indicated by the gray shaded bands. 
Combined, they should be taken as a rough indication of the uncertainty on the
derived rotation curve, but, again, they are not true statistical error
bars. This means that statistical tests, such as evaluating different models
by comparing $\chi^2$ are meaningless and misleading. One should also recall
that the points on the rotation curve are not all independent since the \HI\
velocity field was sampled every half beam-width of the \HI\ observations.  

Figure~\ref{fig:fits} shows that many of the observed
rotation curves in this sample of galaxies exhibit a rapid rise 
followed by a gradual
decline to a constant asymptotic value (as in e.g.\ UGCs~2487, 2953, 3993,
4458). 
This is a fundamental prediction of MOND for galaxies with high central
surface brightness systems.   
Moreover, for these extreme centrally concentrated galaxies, the Newtonian
rotation curves generally coincide with the MOND curve for the inner few kpc
(i.e.\ the mass discrepancy in the inner regions is small).  This is 
particularly evident in UGC 11914 where the entire measured rotation curve
is in the high acceleration regime, and, as MOND would predict, the
discrepancy between the Newtonian and the observed curve is small.

Considering this sample of rotation curves overall we see that in about 10 out 
of the 17 cases the MOND rotation curve closely agrees with the observed
rotation curve. In some cases (e.g.\ UGC~2953) the MOND curve is a spectacular
reproduction of the observed curve. In five other cases (UGCs 3205, 3580, 6786,
9133 and 11670) MOND correctly predicts the general behaviour of the observed
curve but misses details. For two galaxies (UGC 6787, UGC 11852), there are
significant differences between the MOND prediction and the observed rotation
curve, in the overall shape as well as the details.
This is comparable to previous samples (see Sanders \& McGaugh 2002) 
where, in roughly 10\% of the
cases a substantial mismatch between the MOND predictions and the observed
rotation curves was found. 

Given the uncertainties in the derivation of rotation curves and their
precision as a tracer of the radial force distribution, such an occasional
mismatch is not surprising and does not pose an insurmountable problem for
MOND. Nonetheless, we discuss these individual problematic cases and the
possible sources of the disagreement in the following section.

With respect to the implied mass-to-light ratios, the standard interpolating
function (eq.~2) often requires values which are implausibly high, in
particular for the disc (e.g.\ UGCs~3993, 6786). This problem is particularly
apparent when the disc M/L-ratios are compared to those of the bulge: in many
cases the former is larger than the latter (e.g.\ 2953, 3993, 5253, 6786),
whereas naively, one would expect the redder bulge to have a higher M/L. 
With the simpler form given by eq.~3 however, the fitted mass-to-light ratios
of the disc are lower, and usually comparable to or smaller than the bulge
M/L. This indeed suggests that the more gradual transition between Newtonian
and MOND dynamics implied by the \citeauthor{Zhao06} interpolating function
is superior. 

\begin{figure}
 \centerline{\psfig{figure=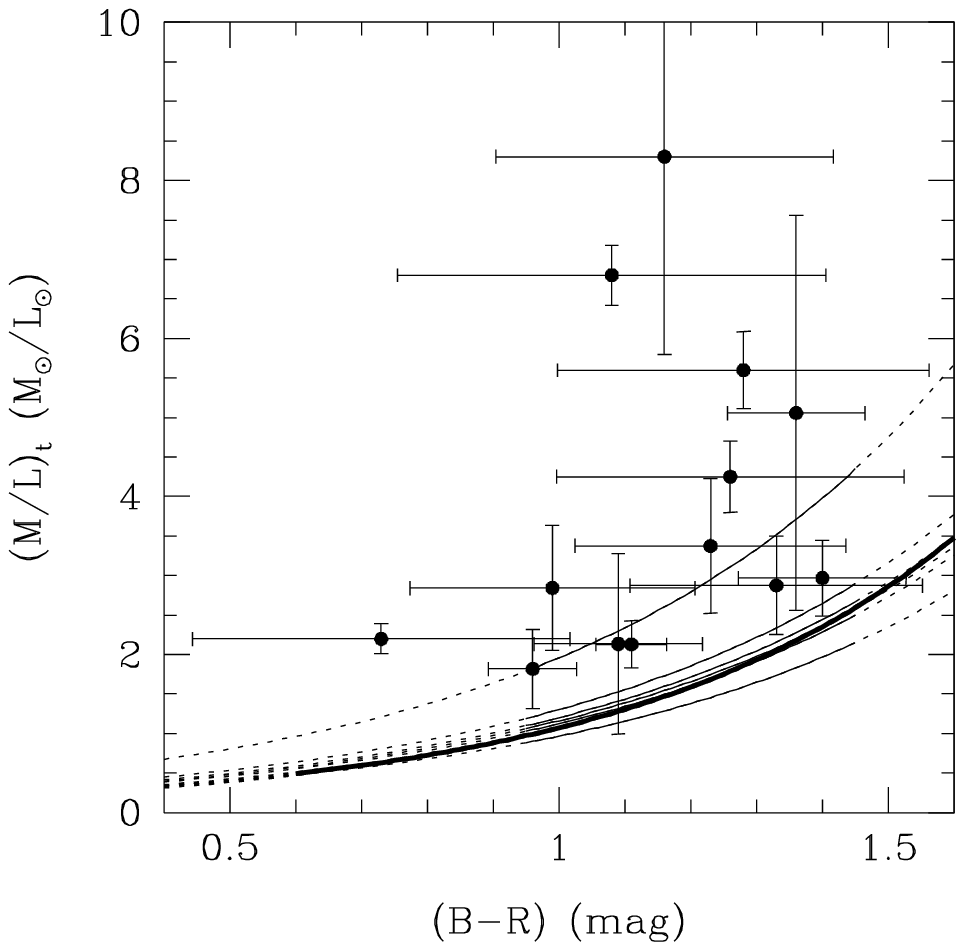,height=8.cm}}
 \caption{Fitted global (bulge+disc) mass-to-light ratios in the R-band vs.\
   B-R colour. Red (gray) and black data points show the mass-to-light ratios
   from the fits with the standard interpolating function $\mu$ and the one
   from \citet{Zhao06} respectively. Errorbars show the photometric errors and 
   the formal fitting errors on M/L. The bold diagonal line shows the relation
   derived by \citet{Bell01}, while the thin lines show models from
   \citet{Portinari04} for different assumptions for the initial stellar mass
   function. The dashed regions show linear extrapolations beyond their bluest
   and reddest models. \label{fig:ML}} 
\end{figure}

In figure~\ref{fig:ML}, we compare the fitted {\em global} (i.e.\ bulge+disc)
mass-to-light ratios (with $\mu$ given by eq.\ 3) to the predictions of recent
population synthesis models of \citet{Bell01} and \citet{Portinari04}.  The
values generally scatter above the models, but if we ignore the two highest
points (UGC 3993 and UGC 11914), the overall trend is similar -- the redder the
galaxy, the higher the M/L. It is also of interest that these two discrepant
objects have the largest uncertainties in inclination. There is a systematic
offset toward higher M/L values than would be predicted by the
\citeauthor{Bell01} models, but, as we see in comparison with the
\citeauthor{Portinari04} models, this can be accommodated by adjustments in
the initial stellar mass function. 

\section{Individual objects}
\label{sec:individualgalaxies}
UGC~3205: The MOND rotation curve deviates significantly from the observed
curve in the inner 10 kpc. This may be a result of streaming motions in the
weak bar in this galaxy, which cause the observed velocities to deviate
significantly from the local circular velocity and lead to asymmetries in the
observed \HI\ velocity field \citepalias{Noordermeer05}. Outside the region
where the velocity field is influenced by the bar, the observed rotation
velocities agree well with the MOND predictions.  

UGC~3580: The MOND rotation curve is generally similar to the observed
curve but the match can be improved considerably 
by increasing the distance to the galaxy by about
  20\%. This is within the uncertainty in the distance to this
relatively nearby object.

UGC 3993:  The overall form of the MOND rotation curve agrees reasonably well
with the observed rotation curve; however, the M/L ratios for the bulge and
disc seem unacceptably large with both the Zhao-Famaey interpolating function
($>8$ for both components) and certainly with the standard function (26 for
the disc). \citet{Edosthesis} notes that this galaxy deviates strongly from
the R-band Tully-Fisher relation in the sense of having a luminosity that is
too low for its rotation velocity. The galaxy is nearly face-on (assumed
inclination of 20 degrees) which means that there is a large uncertainty in
the amplitude of the derived rotation curve. This is a plausible source of the 
anomaly: the galaxy is not under-luminous but the amplitude of the rotation
curve may have been overestimated due to an inclination over-correction. In
this case, the mass-to-light ratios of the bulge and disc will be
correspondingly lower as well. 

UGC~6786:  In the MOND rotation curve shown, the M/L values of the bulge and
disc have been adjusted to match the decline in the observed rotation curve in
the outer regions. In this case, the MOND rotation curve is then too high in
the inner regions. There are large side-to-side differences in the inner
rotation curve -- of about 100 km/s; such asymmetries complicate the
interpretation of the observed radial velocity as a rotation velocity in
the inner regions.  However, at radii larger than 4~kpc the rotation curve is
quite symmetric and the \HI\ and H$\alpha$ data are consistent. 
This is an extreme bulge-dominated galaxy; it appears to be more like an
elliptical with a small disc rather than a normal Sa.  

UGC 6787:  Here the MOND rotation curve differs in form as well as in detail
from the observed rotation curve. This galaxy also deviates strongly from the
R-band Tully-Fisher relation in having too large a rotation velocity for its
luminosity \citep{Edosthesis}. A possible problem here is the distance
estimate. This galaxy lies in the direction of the Ursa Major cluster but has
a redshift 200~km/s larger than the high velocity envelope of this cluster. It
is possible that the true distance is larger than assumed here, but that the
galaxy has a lower apparent Hubble velocity because of in-fall into the UMa
cluster. However, to achieve reasonable agreement with the MOND rotation 
curve, the distance must be increased by almost a factor of two, which seems
implausible. Furthermore, changing the distance does not improve MOND's
ability to reproduce the observed `wiggles' in the rotation curve. An
unmodelled warp beyond 16~kpc could cause the apparent rise in the rotation
curve, but then the inclination of the gas layer in the outer regions would
need to approach $90\deg$ (edge on) to produce the apparent rise of almost
30~km/s. The dip in the rotation curve between 12 and 18~kpc coincides with a
change in the fitted position angle of the tilted rings, indicating deviations
from simple iso-planar gas motion. This does imply that some of the apparent
structure in the rotation curve may be due to the tilted ring fitting
algorithm.

UGC 9133:  Although the general form of the MOND rotation curve agrees with
the observed curve, the detailed structure is not reproduced well. As in the
case of UGC~6787, the fluctuations in the derived rotation curve of this
galaxy are coincident with significant fluctuations in the fitted tilted ring
parameters; in particular there is a shift of more than $10\deg$ in the
position angle of the inclined rings coincident with the bump in the rotation
curve at about 60~kpc. Here the tilted ring program is trying to accommodate a
real systematic twist in the two-dimensional velocity field -- a twist which
may well be due to non-circular motion, in addition to variations in the
orientation of the gas disc. In any case, artifacts in the rotation curve
which are too dependent upon the tilted ring fitting program should be
interpreted with caution. 

UGC 11670:  The inner dip in the rotation curve (at about 4~kpc), which is not
reproduced by MOND, coincides with a fairly dramatic shift in the fitted
position angle of the tilted rings. The true culprit is likely to be
non-circular motion in the potential of the bar. The velocity field is rather
messy in the outer regions; thus, the observational uncertainties on the outer
three points on the rotation curve may be larger than indicated with the
error bars.  The MOND fit to the rotation curve is considerably improved
if the distance is increased by 30\% -- a possibility because
this is the closest galaxy in the sample. 

UGC 11852: This galaxy is clearly warped. There is a large shift in the fitted
inclination of the tilted rings coinciding with the apparent decline in the
rotation curve. Again, it is hard to exclude that possibility that the warp is
accompanied by non-circular motions. In fact, \citetalias{Noordermeer07b}
reported large residual velocities with respect to the tilted ring model,
indicating that some streaming motions are indeed present. A reasonable fit to
this galaxy's rotation curve can also be achieved if it lies at about 75\% of
its adopted Hubble law distance.

\section{Conclusions}
\label{sec:conclusions}
It is rare when an astrophysical hypothesis makes actual predictions in the
realm of extragalactic phenomenology and rarer still when those predictions
are verified. But with respect to the systematic form of galaxy rotation
curves for low surface brightness galaxies, the modified Newtonian dynamics
has been exceptional in this respect. \citeauthor{Milgrom83} predicted in
1983, before a substantial population of low LSB galaxies was discovered, that 
in such objects the rotation curve should rise throughout the visible disc to
the final asymptotic value. Since that time, a population of these objects has
been discovered and observed, and this behaviour is confirmed.  

A supplementary prediction concerned high surface brightness galaxies:  in
these objects the rotation curve should, after a rapid rise in the very
central regions, slowly fall and approach the constant asymptotic value.  

The sample presented here offers the chance to test this prediction. These
early-type galaxies are mostly of high central surface brightness with a
substantial bulge component. The rotation velocities are among the largest
observed, ranging up to 500~km/s.
Such galaxies have been previously difficult to observe because of a relatively
low and patchy gas content, but with the upgraded Westerbork array and
supplementary optical spectroscopic observations in the inner regions, it has
become possible to map two dimensional velocity fields and derive rotation
curves that extend well beyond the bright inner regions. 
The general prediction of MOND with respect to such high surface brightness
objects is confirmed. 

MOND is now seen to successfully account for the systematics and details of
galaxy kinematics from very low rotation, or acceleration, systems
\citep{Milgrom06} to the galaxies presented here with extremely high
rotation velocities, {\it all with the same value of the critical acceleration
parameter $a_0$}. This extends the range of viability of MOND over a range of
a factor of 10 in rotation velocity and $10^4$ in luminosity/mass.  
It therefore becomes increasingly problematic to interpret the success of MOND
as it merely being an algorithm for explaining the relation between dark and
visible matter, as some have suggested. A more natural explanation is to see
MOND as a manifestation of a fundamental modification of gravity or inertia in 
the low acceleration regime -- that MOND is the key to new gravitational
physics. 

\section*{Acknowledgments}
We are grateful to Moti Milgrom and Stacy McGaugh for very helpful
comments on the initial manuscript.

\bibliographystyle{mn2e}
\bibliography{../../references/abbrev,../../references/refs}

\end{document}